\documentclass[conference]{IEEEtran}
\IEEEoverridecommandlockouts

\usepackage{amsmath,amsfonts,amssymb}
\usepackage{graphicx}
\usepackage{cite}
\usepackage{booktabs}
\usepackage{multirow}
\usepackage{bm}
\usepackage{siunitx}
\usepackage{mathtools}
\usepackage[caption=false,font=footnotesize]{subfig}
\usepackage{xcolor}
\usepackage{hyperref}
\usepackage[thinc]{esdiff}
\usepackage{tikz}
\usetikzlibrary{arrows.meta, positioning}

\hypersetup{hidelinks}

\begin{document}

\title{Beyond Path Loss: Altitude-Dependent Spectral Structure Modeling for UAV Measurements
\thanks{This work was supported in part by the National Science Foundation under Grant CNS-2332835 and by the NASA University Leadership Initiative (ULI) under Award 80NSSC25M7102.}
}

\author{
\IEEEauthorblockN{Amir Hossein Fahim Raouf}
\IEEEauthorblockA{
\textit{Department of Electrical and Computer Engineering} \\
\textit{North Carolina State University} \\
Raleigh, NC, USA \\
amirh.fraouf@ieee.org}
\and
\IEEEauthorblockN{\.{I}smail G\"{u}ven\c{c}}
\IEEEauthorblockA{
\textit{Department of Electrical and Computer Engineering} \\
\textit{North Carolina State University} \\
Raleigh, NC, USA \\
iguvenc@ncsu.edu}
}

\maketitle

\begin{abstract}
This paper presents a measurement-based framework for characterizing altitude-dependent spectral behavior of signals received by a tethered Helikite unmanned aerial vehicle~(UAV). Using a multi-year spectrum measurement campaign in an outdoor urban environment, power spectral density snapshots are collected over the 89~MHz--6~GHz range. Three altitude-dependent spectral metrics are extracted: band-average power, spectral entropy, and spectral sparsity.
We introduce the Altitude-Dependent Spectral Structure Model~(ADSSM) to characterize the spectral power and entropy using first-order altitude-domain differential equations, and spectral sparsity using a logistic function, yielding closed-form expressions with physically consistent asymptotic behavior. The model is fitted to altitude-binned measurements from three annual campaigns at the AERPAW testbed across six licensed and unlicensed sub-6 GHz bands.
Across all bands and years, the ADSSM achieves low root-mean-square error and high coefficients of determination. Results indicate that power transitions occur over narrow low-altitude regions, while entropy and sparsity evolve over broader, band-dependent altitude ranges, demonstrating that altitude-dependent spectrum behavior is inherently multidimensional. By explicitly modeling altitude-dependent transitions in spectral structure beyond received power, the proposed framework enables spectrum-aware UAV sensing and band selection decisions that are not achievable with conventional power- or threshold-based occupancy models.
\end{abstract}
\begin{IEEEkeywords}
AERPAW, altitude-dependent channel modeling, cellular networks, radio maps, sparsity, spectral entropy, spectrum measurements, UAV communications.
\end{IEEEkeywords}

\section{Introduction}
Uncrewed Aircraft Systems~(UAS) play an increasingly critical role in mission-critical and commercial applications, including command and control~(C2), payload data delivery, and situational awareness.~\cite{FotouhiUAVCellularSurvey2019,geraci2022will}. Relative to terrestrial user equipment~(UE), UAS operate under fundamentally different propagation conditions, characterized by higher line-of-sight~(LoS) probability, increased interference from multiple base stations~(BSs), and broader coverage footprints~\cite{khuwaja2018survey}. Accurate characterization of altitude-dependent signal behavior is therefore essential for effective interference management, airspace coexistence, and reliable UAS network design~\cite{raouf2025wireless}.

Most prior work on UAV communications has focused on path loss, shadowing, and fast fading as functions of altitude and horizontal distance, typically within a single-band single-operator context, see e.g.,~\cite{AlHouraniA2GPathLoss2014,willink2015measurement}. In contrast, practical UAV systems operate in spectrally dense environments where multiple bands and operators coexist and where the spectral structure itself, that is the distribution of power across frequency, varies with altitude. For tasks such as band selection, interference-aware trajectory planning, or RF-based sensing, received power alone is insufficient. It is necessary to characterize how spectral structure and spectrum occupancy evolve with altitude.

Existing spectrum occupancy studies commonly rely on scalar descriptors such as duty cycle, occupancy probability above a fixed threshold, or median and percentile power levels to summarize band usage~\cite{yucek2009survey,wellens2007evaluation,LopezBenitezMethodology2010,ChantaveerodFMOccupancy2021,KozlowskiISMOccupancy2021}. While effective for quantifying average utilization, these metrics collapse the internal frequency-domain structure of the received signal and do not capture how power is distributed across subcarriers or how this distribution evolves with altitude. In particular, duty cycle and occupancy probability are highly sensitive to threshold selection and provide no information on spectral dispersion or fragmentation, while median power conflates changes in aggregate power with changes in spectral shape. Consequently, such metrics cannot distinguish between bands with similar average power but fundamentally different spectral organization, nor can they reveal altitude-dependent transitions in interference structure once band-average power has saturated.

This paper addresses the gap between altitude-dependent air-to-ground~(A2G) channel modeling and spectrum occupancy characterization by:
\begin{itemize}
    \item conducting a multi-year multi-band spectrum measurement campaign using a Helikite platform in an outdoor urban environment;
    \item defining three spectral metrics that jointly capture amplitude structure and occupancy within licensed and shared bands (e.g., FM broadcast, cellular downlink, ISM);
    \item introducing the Altitude-Dependent Spectral Structure Model~(ADSSM), which extends power-only A2G models by explicitly capturing altitude-dependent changes in spectral structure;
    \item validating ADSSM across multiple bands and campaign years.
\end{itemize}

The resulting framework complements classical path-loss-only models by providing a compact and physically interpretable description of how the statistical structure of the received spectrum evolves with altitude. 


\section{Measurement Campaign and Dataset}\label{sec:measurement_process}

The measurements were obtained using the AERPAW spectrum-monitoring experiment, deployed on a tethered Helikite platform during the Packapalooza festival on the NC State University campus over multiple years. The goal of the campaign was to capture sub-6~GHz spectrum activity in a dense urban environment under realistic network load conditions.

\begin{figure}[!t]
	\centering
        \includegraphics[width=\linewidth]{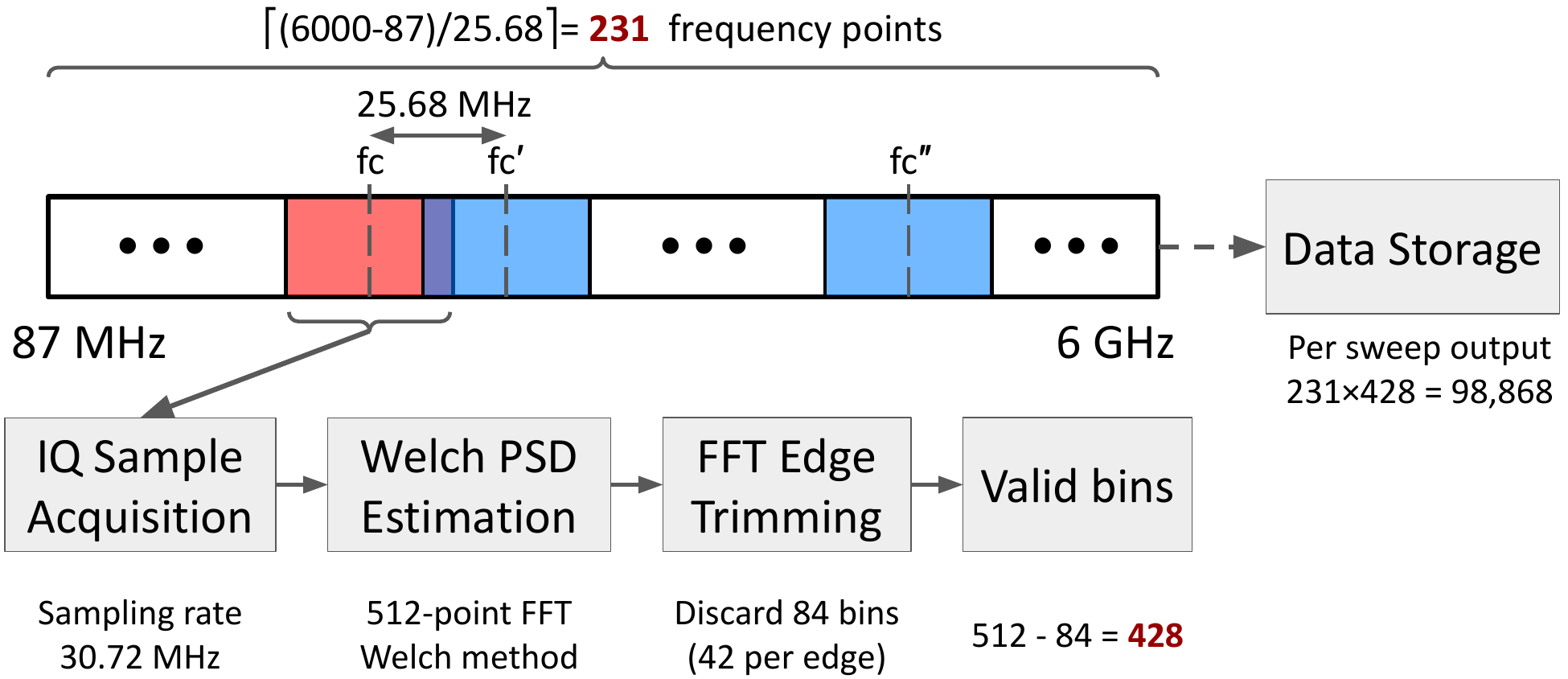}\vspace{-2mm}
	\caption{Spectrum sweep procedure for spectrum data collection using a helikite-mounted portable node.}\label{fig:sweep_procedure_AERPAW}
\end{figure}

As the measurement campaign and associated dataset are described in~\cite{maeng2025altitude, raouf2025wireless}, only a brief overview of the measurement procedure and dataset is provided here.
The payload consisted of a USRP-based software-defined radio~(SDR) and a GNSS module providing altitude and time synchronization. The spectrum-monitoring experiment operates by sweeping center frequencies from 87~MHz to 6~GHz in a loop. For each center frequency, the USRP collects $500$k IQ samples at a sampling rate of 30.72~MHz, and Welch’s method is applied using a 512-point FFT. To suppress edge artifacts, FFT bins at the band edges are discarded, and the center frequency is incremented by 25.68~MHz across the sweep. Each sweep is stored for post-processing, and Fig.~\ref{fig:sweep_procedure_AERPAW} illustrates the spectrum sweep procedure for the helikite-mounted platform.
Although experimental parameters such as USRP gain, sampling rate, number of sweep points, and FFT size are configurable, they were held constant across measurement campaigns to ensure cross-year comparability, with only minor adjustments to experiment duration imposed by operational constraints.


The Helikite altitude was managed through a ground-based tether. Wind and operational constraints caused lateral drift and non-uniform altitude trajectories across years. 
Throughout all experiments, the SDR operated passively and did not modify cellular network configurations such as carrier aggregation, scheduling, or base-station transmission behavior. The resulting dataset therefore represents an unaltered view of real-world spectrum usage and is suitable for altitude-dependent analysis of power, spectral entropy, and sparsity across multiple commercial cellular bands and unlicensed allocations.

\looseness=-1
The raw PSD output from each sweep is mapped onto a global frequency grid indexed by frequency bins. For each band $b$, a contiguous set of bins $\mathcal{F}_b$ is identified according to 3GPP and FCC allocations. For each band and sweep, the corresponding PSD samples are extracted to obtain $P_b(f,h)$ for all $f \in \mathcal{F}_b$ at instantaneous altitude $h$. 
To mitigate small-scale variability and account for nonuniform altitude sampling, the altitude range is partitioned into uniform bins of size $\Delta h$ (e.g., 10~m). For each bin $k$ with center altitude $h_{c,k}$, all PSD samples within the bin are aggregated, and bin-level statistics are computed for each spectral metric, as detailed in Section~\ref{sec:spectral_metrics}.

\section{Spectral Metrics and Their Physical Meaning}\label{sec:spectral_metrics}

Let $P_{b}(f,h)$ denote the linear-scale measured power at frequency bin~\footnote{Throughout this section, the term ``frequency bin'' refers to receiver-defined FFT bins obtained from passive PSD measurements, without implying knowledge of transmitter subcarriers or resource allocation.} $f\in \mathcal{F}_b$ and altitude $h$ for band~$b$. We consider three complementary metrics: band-average power, spectral entropy, and spectral sparsity.

\subsection{Band-Average Power}

The band-average power at altitude $h$ is defined as
\begin{equation}
    P_b(h) = 10 \log_{10} \left( \frac{1}{|\mathcal{F}_b|} 
    \sum_{f \in \mathcal{F}_b} P_{b}(f,h)\right)\,\text{dB}.
\end{equation}

This is the conventional band-integrated power, capturing the aggregate measured power from all BSs contributing to band~$b$. 
As altitude increases, $P_b(h)$ is expected to exhibit a transition from clutter-limited to LoS-dominated behavior, with a corresponding increase in average received power and decrease in variability~\cite{maeng2025altitude}.

\subsection{Spectral Entropy}

We define the normalized power distribution within band $b$ at altitude $h$ as
\begin{equation}
    p_b(f,h) = \frac{P_{b}(f,h)}{\sum_{f' \in \mathcal{F}_b} P_{b}(f',h) + \varepsilon},
\end{equation}
where $\varepsilon$ is a small positive constant to avoid division by zero. The corresponding Shannon spectral entropy is
\begin{equation}
    H_b(h) = - \sum_{f \in \mathcal{F}_b} p_b(f,h) \log_2 p_b(f,h)\,\text{bits}.
\end{equation}

Spectral entropy quantifies the distribution of received power across frequency within a given band, as observed through receiver-defined frequency bins. When the in-band power is concentrated in a limited set of dominant spectral components, the entropy is low, whereas a more uniformly distributed spectrum across frequency bins results in higher entropy. In OFDM-based systems, variations in spectral entropy can qualitatively reflect changes in the apparent spectral occupancy and aggregate interference at the receiver. However, since the measurements are based on passive PSD snapshots without synchronization, decoding, or control-channel awareness, spectral entropy must be interpreted strictly as an observational descriptor of the received spectral shape, rather than as a direct measure of traffic load, subcarrier allocation, or MAC-layer scheduling. Unlike band-average power, which is primarily influenced by large-scale path loss, spectral entropy captures how the internal spectral structure within a band evolves with altitude.

\subsection{Spectral Sparsity}
\looseness=-1
Sparsity quantifies the fraction of frequency bins that contain detectable signal energy. Following common practice in spectrum occupancy studies~\cite{LopezBenitezMethodology2010, islam2008spectrum}, we define a band-specific detection threshold $\gamma_b = P_{b,\text{noise}} + \Delta_{\text{th}}$, where $P_{b,\text{noise}}$ is an empirical noise floor estimate based on the fifth percentile of the band's PSD samples over the entire campaign, and $\Delta_{\text{th}}$ (e.g., 3~dB) is a conservative margin above noise~\cite{wellens2007evaluation}. This threshold is fixed per band and campaign year rather than per altitude to avoid confounding sparsity variations with altitude-dependent noise estimation. Similar percentile-based thresholding approaches have been adopted in recent spectrum occupancy studies, for example in FM band measurements~\cite{ChantaveerodFMOccupancy2021}.

The sparsity metric at altitude $h$ is then calculated as
\begin{equation}
    S_b(h) = \frac{1}{|\mathcal{F}_b|} \sum_{f\in \mathcal{F}_b} 
    \mathbf{1} \bigl\{ P_b(f,h) > \gamma_b \bigr\},
\end{equation}
where $\mathbf{1}\{\cdot\}$ is the indicator function. Values of $S_b(h)$ near zero indicate that few bins are above threshold, whereas values near one indicate that most of the band is occupied. Sparsity thus captures band occupancy and visibility of transmitters in a manner complementary to power and entropy.

\subsection{Transition Regions and 10–90\% Metrics}
\looseness=-1
For each altitude-dependent metric, we focus on the altitudes at which meaningful transitions occur, namely where behavior shifts from clutter dominated to LoS dominated. To quantify these transitions, we define for each band and metric the 10\%, 50\%, and 90\% transition heights $(h_{10}, h_{50}, h_{90})$ as the altitudes at which the metric reaches 10\%, 50\%, and 90\% of its total change between the low-altitude value and the asymptotic high-altitude limit.
This definition offers several advantages:
\begin{itemize}
    \item It is \emph{scale invariant} and applies consistently to power in dB, sparsity on the unit interval, and entropy in bits, enabling direct comparison across bands with different dynamic ranges.
    \item It robustly captures the effective clutter region associated with rooftops and tree canopies and the altitude range over which LoS probability increases rapidly~\cite{AlHouraniA2GPathLoss2014,khuwaja2018survey}.
    \item The midpoint $h_{50}$ corresponds to the altitude at which the metric slope with respect to height is maximized, analogous to a breakpoint or corner frequency, and is therefore well suited for engineering design.
\end{itemize}

In Section~\ref{sec:validation}, these transition heights are computed from the fitted ADSSM curves and interpreted across different frequency bands.

\section{ADSSM Model}

The ADSSM characterizes the altitude evolution of $P_b(h)$, $H_b(h)$, and $S_b(h)$ for each band $b$.

\subsection{First-Order Differential Model for Power and Entropy}

Let $X_b(h)$ denote an altitude-dependent spectral metric for band $b$, representing either band-average power or spectral entropy. The evolution of $X_b(h)$ is modeled by the first-order linear differential equation
\begin{equation}
    \diff{X_b(h)}{h}= \frac{X_{b}(\infty) - X_b(h)}{\Delta h_b},
    \label{eq:ode_generic}
\end{equation}
where $X_{b}(\infty)$ denotes the asymptotic high-altitude value and $\Delta h_b>0$ is a characteristic altitude constant. This model assumes that the rate of change with altitude is proportional to the remaining distance from the limiting value, resulting in gradual convergence toward a LoS-dominated regime.

The first-order structure is motivated by the physical mechanisms governing air-to-ground propagation. In cluttered urban and suburban environments, the most significant changes in received signal characteristics occur as the platform ascends through the altitude range in which the LoS probability increases rapidly. Empirical models show that both LoS probability and excess path loss approach their limiting values approximately exponentially with altitude~\cite{AlHouraniA2GPathLoss2014}. Spectral metrics derived from the received signal, including band-average power and spectral entropy, therefore exhibit similar altitude-dependent behavior. Once the platform rises above the effective clutter height, further altitude increases yield diminishing variations, which naturally motivates a first-order relaxation model. The model in~\eqref{eq:ode_generic} is intended to capture the dominant monotonic altitude trends over the measured altitude range and does not account for potential power degradation at very high altitudes, where visibility and path loss effects may reverse the trend.

Integrating~\eqref{eq:ode_generic} with the initial condition $X_b(0)$ yields
\begin{equation}
    X_b(h) = X_{b}(\infty)
    - \bigl(X_{b}(\infty) - X_{b}(0)\bigr) e^{-h/\Delta h_b},
    \label{eq:generic_solution}
\end{equation}
where $X_b(0)$ denotes the effective ground-level value. Applying~\eqref{eq:generic_solution} to band-average power yields
\begin{equation}
    P_b(h) = P_{b}(\infty)
    - \bigl(P_{b}(\infty) - P_{b}(0)\bigr) e^{-h/h_{c,b}},
\end{equation}
where $h_{c,b}$ represents the clutter height governing the transition from clutter-limited to LoS-dominated propagation.

Similarly, the spectral entropy is modeled as
\begin{equation}
    H_b(h) = H_{b}(\infty)
    - \bigl(H_{b}(\infty) - H_{b}(0)\bigr) e^{-h/h_{e,b}},
\end{equation}
\looseness=-1 where $h_{e,b}$ controls the rate at which the spectral structure stabilizes with altitude. To enable fair comparison across frequency bands with different bandwidths and spectral resolutions, the spectral entropy is normalized by its maximum value $\log_2 |\mathcal{F}_b|$, yielding $\tilde{H}_b(h) \in [0,1]$.
Although spectral entropy typically evolves more slowly with altitude than band-average power, the proposed first-order model captures its monotonic convergence toward a limiting spectral regime at higher altitudes. Together, these exponential models provide a compact and physically interpretable description of altitude-dependent spectral evolution across frequency bands and measurement campaigns.

\subsection{Logistic Model for Sparsity}
Sparsity is modeled using a logistic activation function
\begin{equation}
    S_b(h) = \frac{1}{1 + \exp\bigl(-k_b (h - h_{s,b})\bigr)},
\end{equation}
where $k_b>0$ controls the transition steepness and $h_{s,b}$ denotes the altitude at which $S_b(h)=0.5$. This form is motivated by the progressive activation of additional base stations and sub-bands as LoS conditions emerge with increasing altitude and by the saturation of sparsity at high altitudes when most sub-bands become observable. Empirically, sparsity exhibits sigmoidal transitions rather than purely exponential behavior, which is well captured by the logistic model. In addition, this formulation guarantees that $S_b(h)$ remains within the unit interval for all altitudes, preserving physical interpretability and enabling a natural definition of 10--90\% transition heights.

\subsection{Parameter Estimation and Benchmark Models}

For each year and band, we fit the parameters of the ADSSM model to the altitude-binned means, $\{P_b(h_{c,k})\}$, $\{H_b(h_{c,k})\}$, and $\{S_b(h_{c,k})\}$, defined in Section~\ref{sec:measurement_process}. Let $\bm{\theta}_P = [P_{b}(\infty), P_{b}(0), h_{c,b}]$, $\bm{\theta}_H = [H_{b}(\infty), H_{b}(0), h_{e,b}]$, and $\bm{\theta}_S = [k_b, h_{s,b}]$ denote the parameter vectors. We solve
\begin{equation}
    \bm{\theta}_P^\star = \arg\min_{\bm{\theta}_P} 
    \sum_k \bigl( P_b(h_{c,k}) - \hat{P}_b(h_{c,k}; \bm{\theta}_P) \bigr)^2,
\end{equation}
and similarly for entropy and sparsity, where $\hat{P}_b(\cdot)$ is the ADSSM prediction.
We use unconstrained non-linear least squares implemented via a derivative-free search (e.g., Nelder--Mead) with carefully initialized starting points based on the first and last altitude bins. For entropy, where dynamic range may be small, we allow a reduced-parameter version in which $h_{e,b}$ is fixed to a characteristic value derived from the altitude span, and only $H_{b}(\infty)$ and $H_{b}(0)$ are estimated. This mitigates identifiability issues when entropy is nearly flat.

\begin{figure*}[!t]
\centering
\subfloat[FM broadcast]{%
\includegraphics[width=0.31\textwidth]{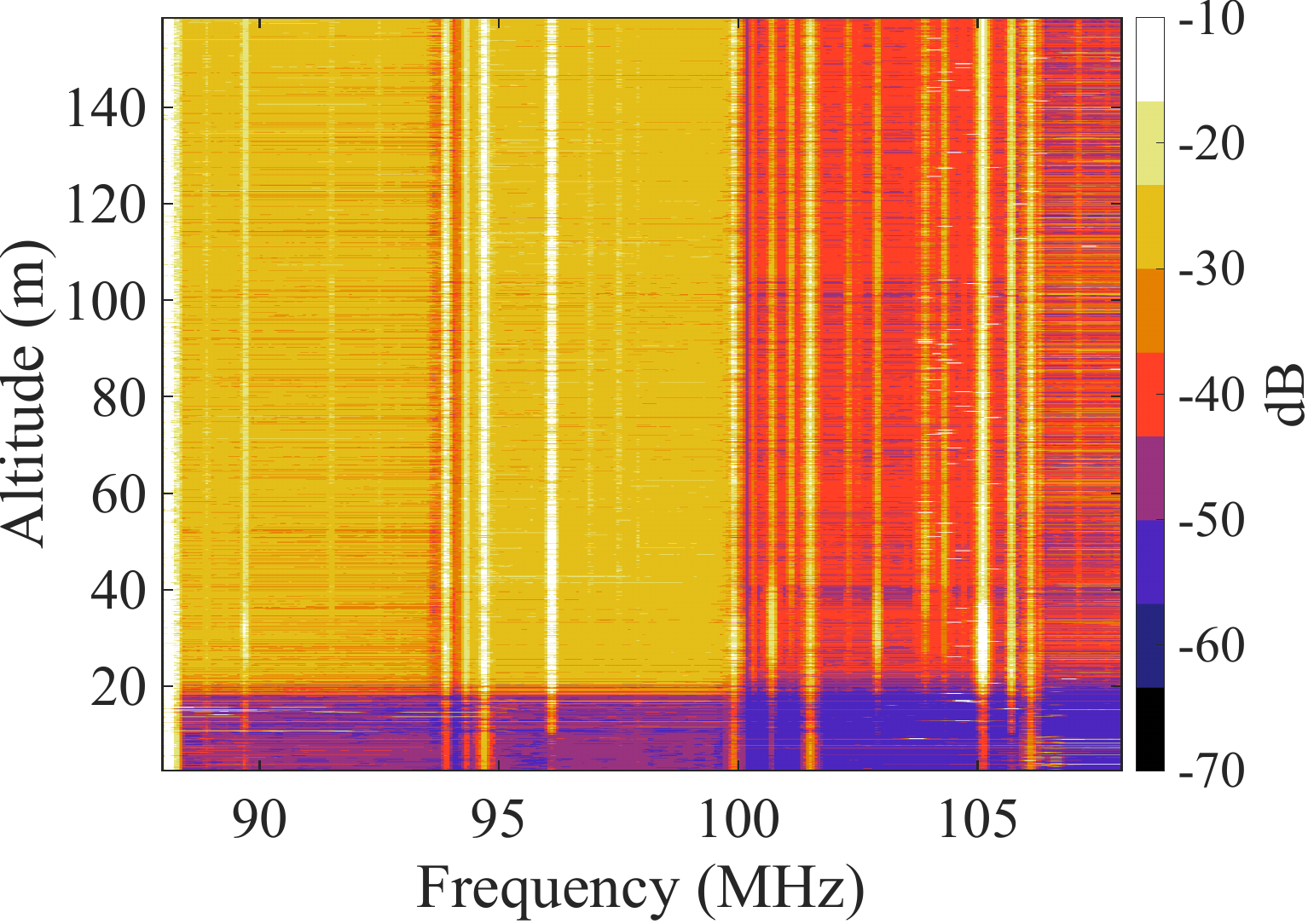}
\label{fig:F0_AltFreqHeatmap_2025_FM}}
\hfill
\subfloat[5G NR n71 downlink]{%
\includegraphics[width=0.31\textwidth]{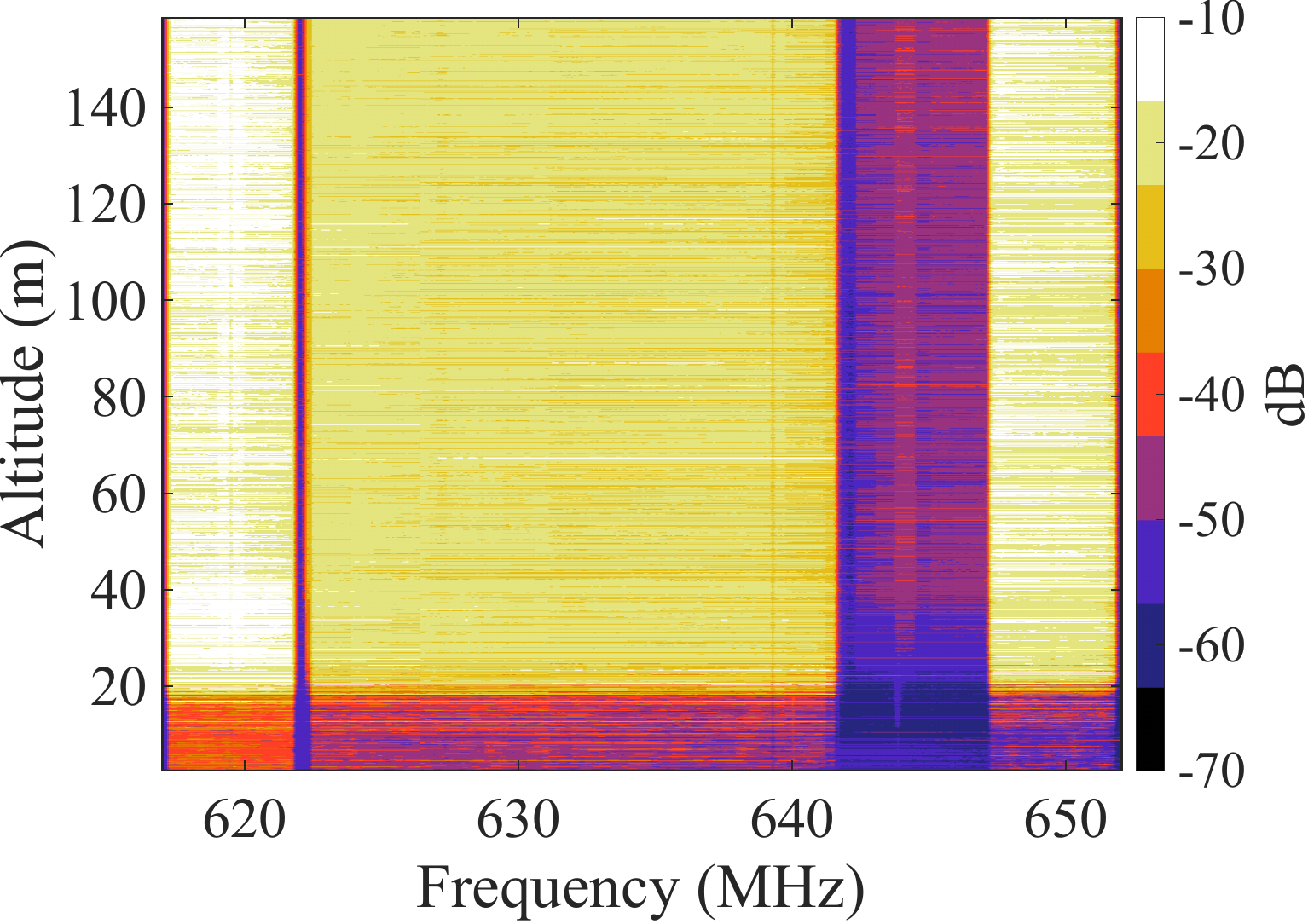}
\label{fig:F0_AltFreqHeatmap_2025_5G_Band_n71_DL}}
\hfill
\subfloat[LTE Band~13 downlink]{%
\includegraphics[width=0.31\textwidth]{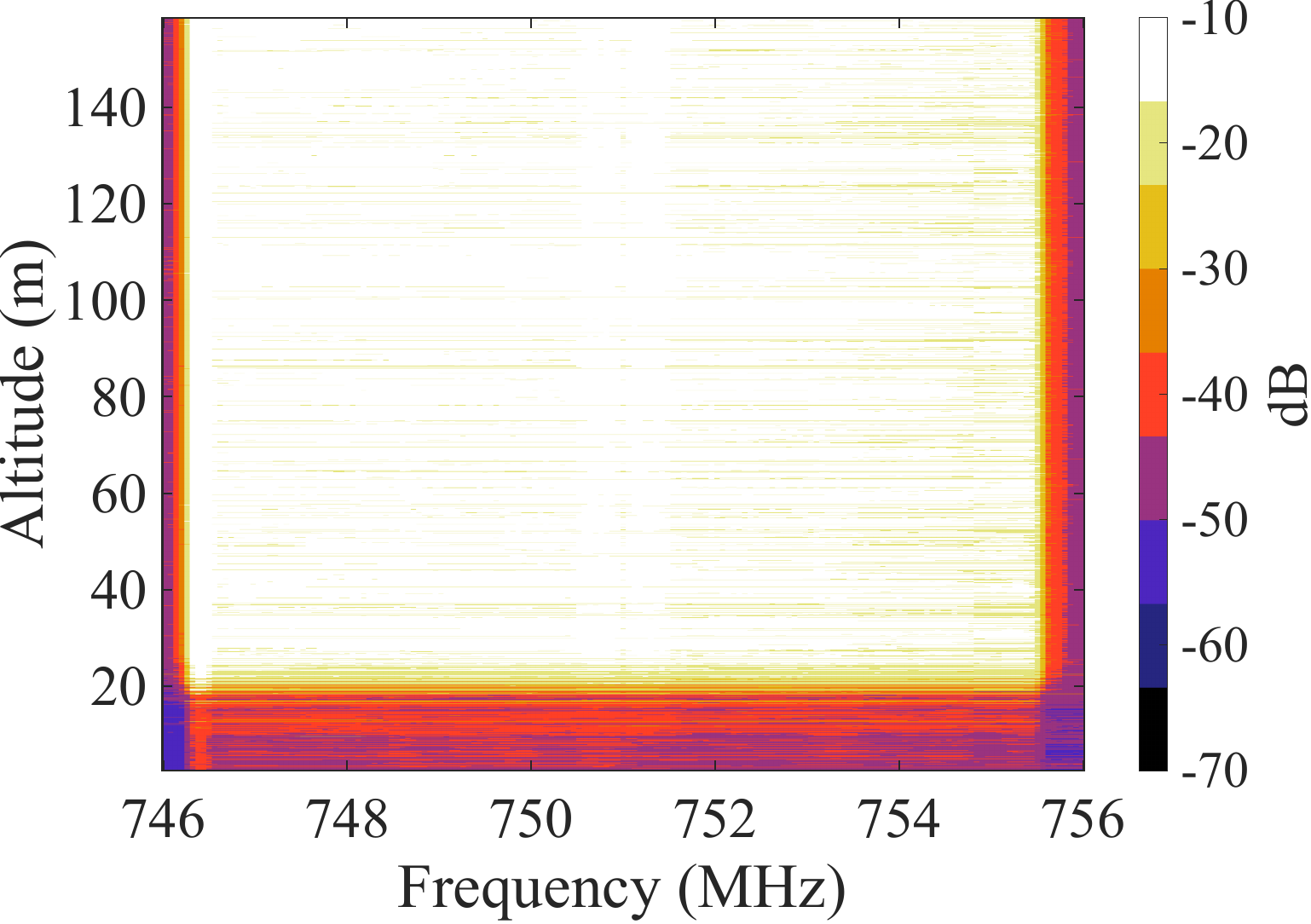}
\label{fig:F0_AltFreqHeatmap_2025_LTE_Band_13_DL}} \\

\subfloat[ISM band]{%
\includegraphics[width=0.31\textwidth]{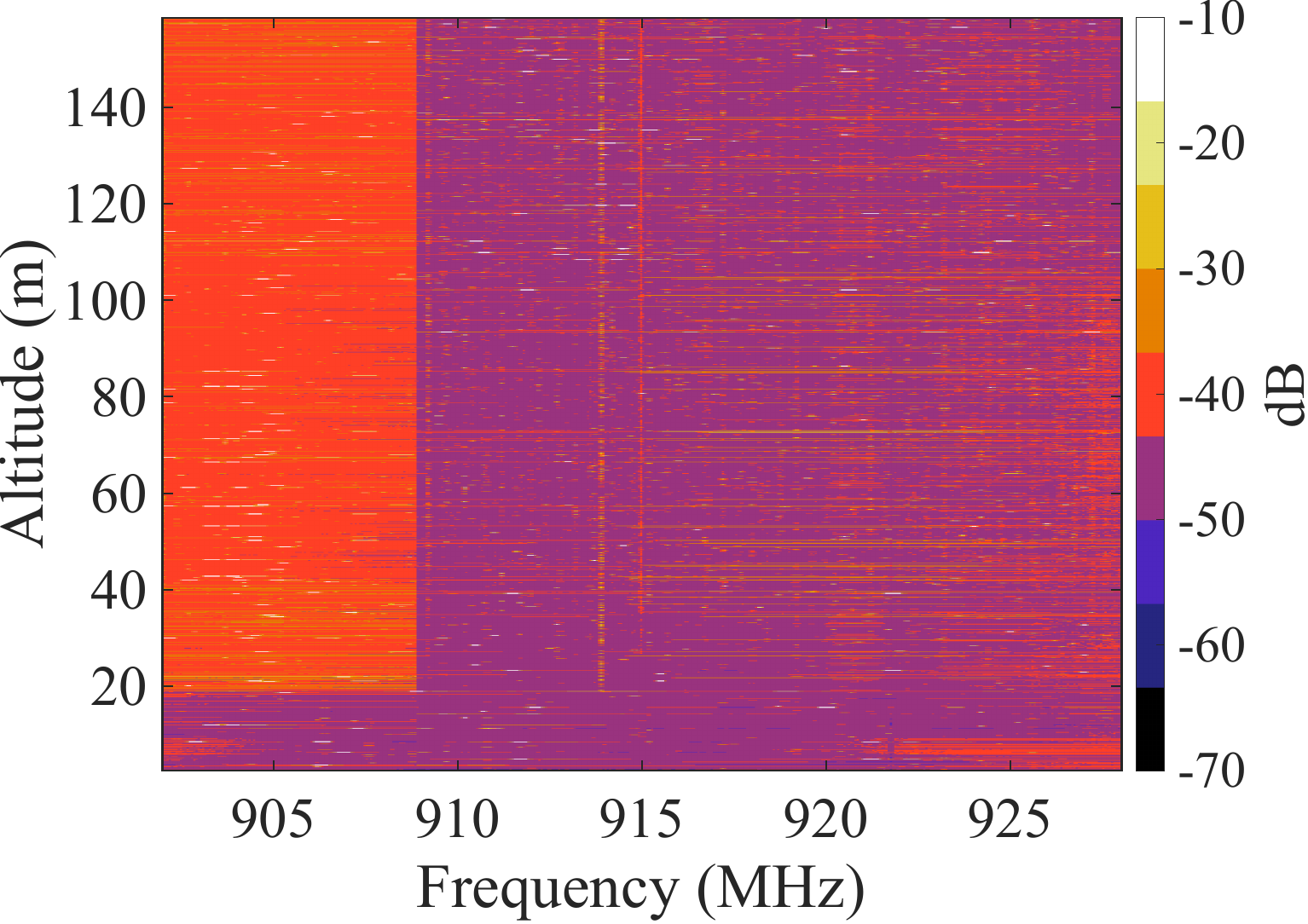}
\label{fig:F0_AltFreqHeatmap_2025_ISM}}
\hfill
\subfloat[CBRS]{%
\includegraphics[width=0.31\textwidth]{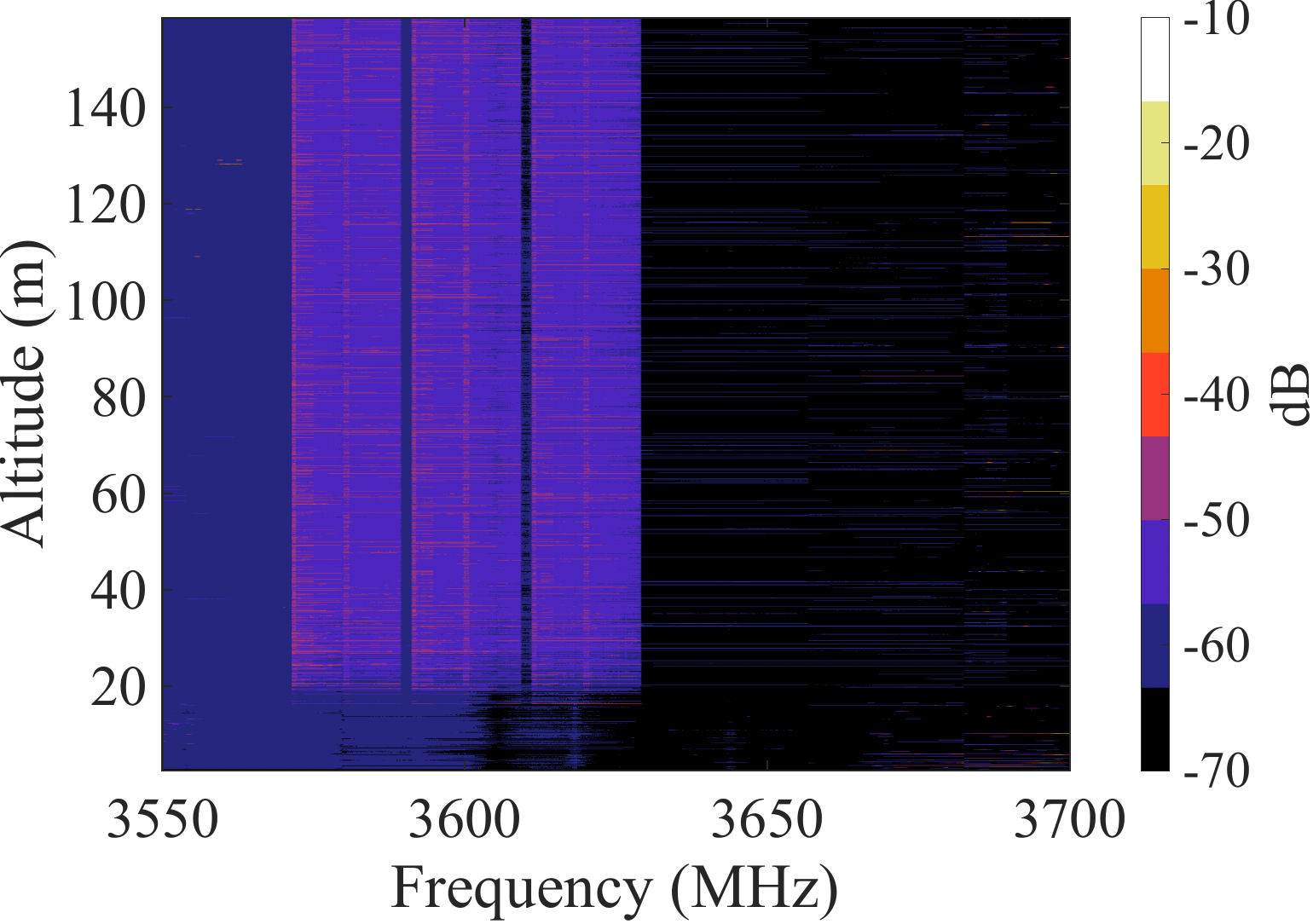}
\label{fig:F0_AltFreqHeatmap_2025_CBRS}}
\hfill
\subfloat[5G NR C-band]{%
\includegraphics[width=0.31\textwidth]{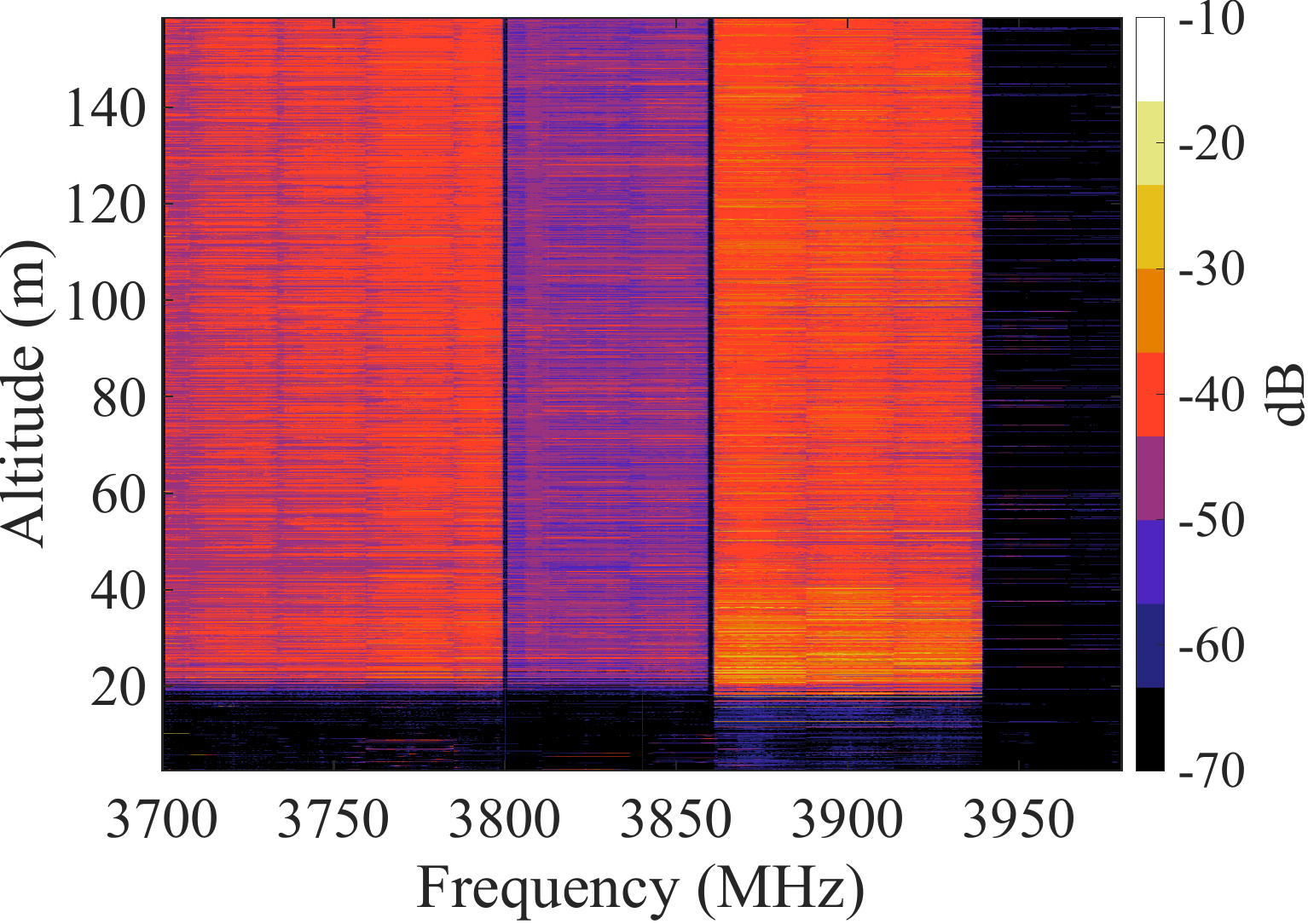}
\label{fig:F0_AltFreqHeatmap_2025_C-Band}}

\caption{Altitude--frequency heatmaps for six representative spectrum allocations:
(a) FM broadcast,
(b) 5G NR n71 downlink,
(c) LTE Band~13 downlink,
(d) ISM band,
(e) CBRS, and
(f) 5G NR C-band.
These raw measurements provide the foundational altitude-dependent spectral structure from which subsequent power, entropy, and sparsity metrics are derived.}
\label{fig:altfreq_updatedbands}
\end{figure*}

\section{Empirical Characterization and ADSSM Validation}\label{sec:validation}

This section presents the empirical results for the three spectral metrics and evaluates the ADSSM model across all frequency bands and measurement campaigns. We consider six representative U.S. frequency bands: FM broadcast (88–108 MHz), 5G NR n71 downlink (617–652 MHz), LTE Band 13 downlink (746–756 MHz), the ISM band (902–928 MHz), the CBRS band (3.55–3.7 GHz), and the 5G NR C-band downlink (3.7–3.98 GHz). Owing to page limitations, results are shown only for the Packapalooza 2025 campaign. Nevertheless, across all evaluated years and bands, the ADSSM closely tracks power, entropy, and sparsity while achieving low RMSE and high $R^2$ values.

\subsection{Altitude--Frequency Heatmaps}

The altitude--frequency heatmaps in Fig.~\ref{fig:altfreq_updatedbands} summarize the raw spectral structure for the six bands under consideration and reveal clear altitude-dependent patterns that differ markedly across frequency ranges and service types. In the FM broadcast band, strong narrowband carriers are visible across the entire altitude range, with a pronounced transition near low altitudes where clutter attenuation suppresses weaker stations. As altitude increases, additional FM carriers emerge and the overall band becomes more uniformly occupied, reflecting increased LoS visibility to geographically distributed transmitters. Similar but more structured behavior is observed in the cellular downlink bands. For 5G~NR~n71 and LTE~Band~13, the heatmaps indicate relatively uniform received power at higher altitudes, while lower altitudes exhibit reduced power and increased frequency-selective behavior due to shadowing and blockage. The ISM band exhibits a heterogeneous pattern with intermittent narrowband activity and comparatively weak altitude dependence, consistent with unlicensed low-power emitters and sporadic interference sources. In the CBRS and 5G NR C-band allocations, distinct sub-band structures and guard bands are clearly visible, with received power increasing rapidly over a limited altitude range before saturating at higher altitudes. Across all bands, the heatmaps illustrate a common transition from clutter-limited conditions at low altitude to a LoS-dominated regime at higher altitude, while also highlighting substantial differences in spectral occupancy and structure that motivate the joint use of power, entropy, and sparsity metrics in the proposed ADSSM framework.

\begin{figure*}[!t]
    \centering
    \subfloat[Band-average power vs. altitude]{
        \includegraphics[width=0.31\linewidth]{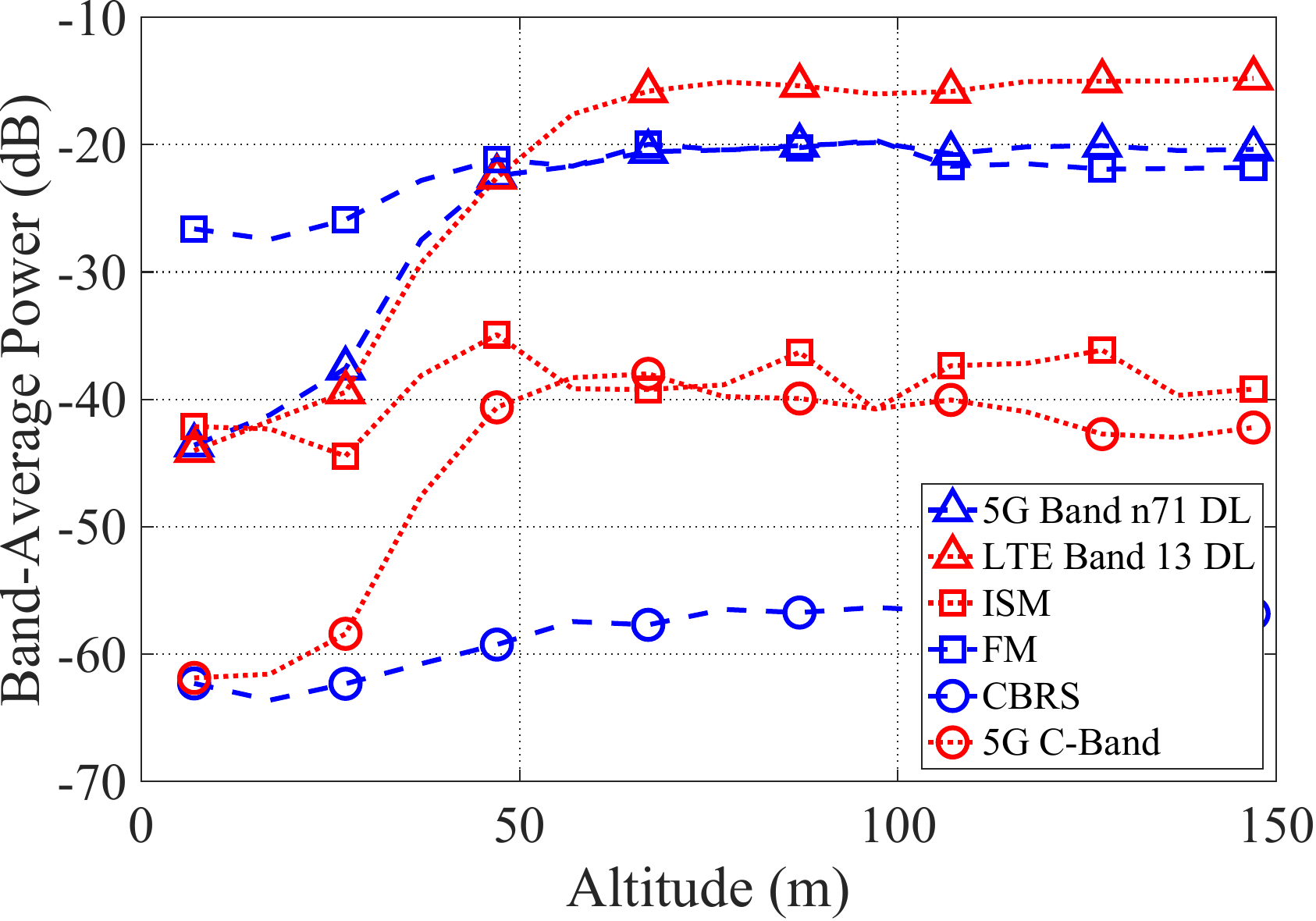}
        \label{fig:F1_Power_vs_Altitude_2025}} 
    \subfloat[Normalized spectral entropy vs. altitude]{
        \includegraphics[width=0.31\linewidth]{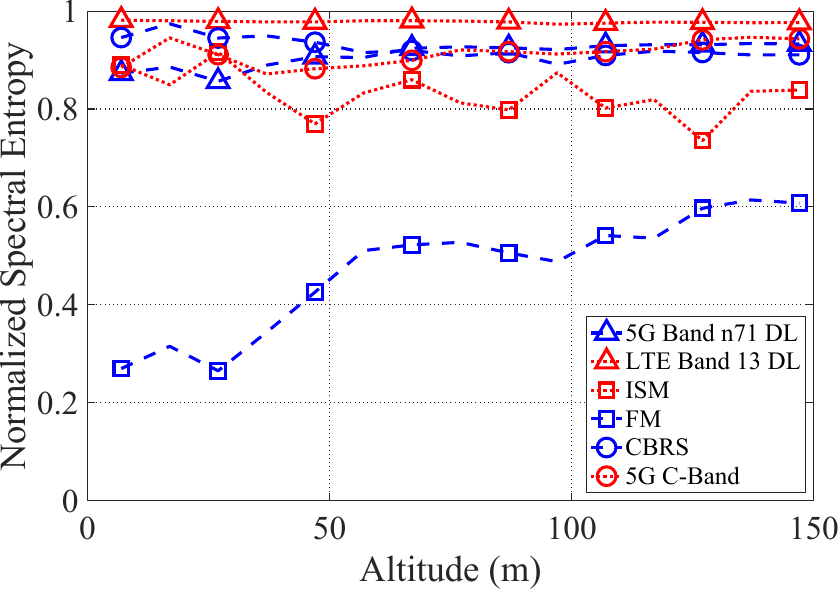}
        \label{fig:F1_Entropy_vs_Altitude_2025}} 
    \subfloat[Band-average sparsity vs. altitude]{
        \includegraphics[width=0.31\linewidth]{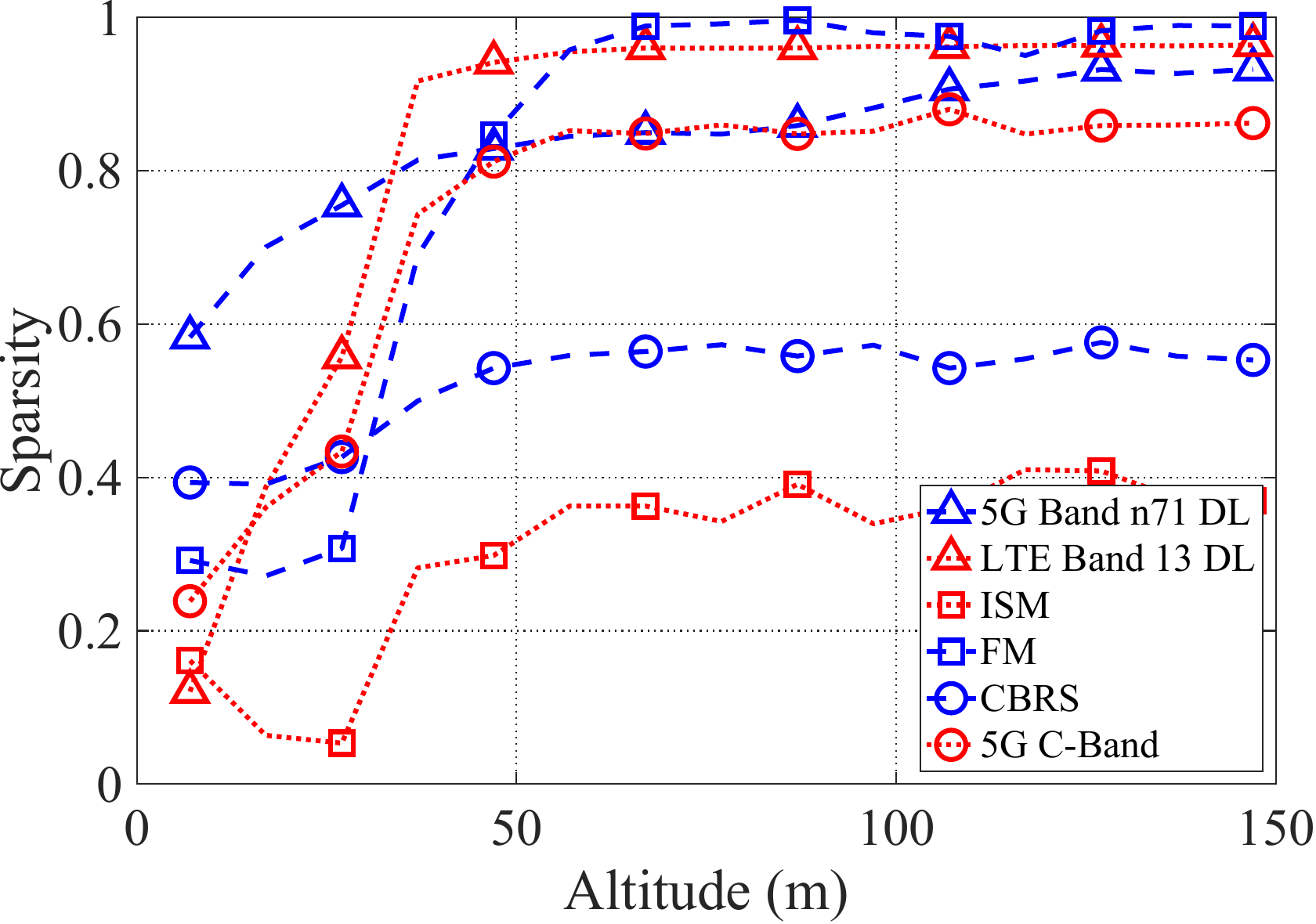}
        \label{fig:F1_Sparsity_vs_Altitude_2025}} 
    \caption{Altitude-dependent band-average metrics for six measured allocations:
    (a) band-average power, (b) normalized spectral entropy, and (c) sparsity.
    These metrics aggregate the altitude–frequency structure in Fig.~\ref{fig:altfreq_updatedbands} and quantify each band’s transition from clutter-limited propagation at low altitudes to its high-altitude asymptotic regime.
    }
    \label{fig:F1_metrics_allbands}
\end{figure*}

\subsection{Altitude-Dependent Metrics}

The altitude-dependent band-average metrics in Fig.~\ref{fig:F1_metrics_allbands} summarize the aggregate spectral behavior across the six bands under consideration and provide a compact view of how each allocation transitions with height. The band-average power curves in Fig.~\ref{fig:F1_Power_vs_Altitude_2025} exhibit a clear increase with altitude for all cellular bands, with the most pronounced gains occurring over a limited low-altitude range before saturating at higher altitudes. This behavior reflects the transition from clutter-limited propagation to a LoS-dominated regime. In contrast, the FM broadcast band shows relatively high power even at low altitudes, consistent with high-power transmitters and elevated antenna heights, while the ISM and CBRS bands remain power limited across the full altitude range due to lower transmit powers and heterogeneous deployments.

The normalized spectral entropy trends in Fig.~\ref{fig:F1_Entropy_vs_Altitude_2025} further highlight differences in spectral structure across frequency bands. For FM broadcast, entropy increases with altitude as additional carriers become observable, then gradually stabilizes, reflecting a transition from sparse to more distributed spectral occupancy. Cellular bands exhibit consistently high normalized entropy with only mild altitude-dependent variations, indicating that their in-band spectral structure remains largely uniform as altitude increases. The ISM band shows comparatively larger fluctuations in entropy, consistent with intermittent and spatially localized emitters. 

The sparsity curves in Fig.~\ref{fig:F1_Sparsity_vs_Altitude_2025} capture complementary behavior. FM and cellular bands exhibit strong sigmoidal transitions, with sparsity increasing rapidly over a narrow altitude range and approaching saturation at higher altitudes as more transmitters and sub-bands enter LoS. The ISM band remains relatively sparse across all altitudes, while CBRS exhibits intermediate sparsity levels due to partial utilization and regulatory constraints. Collectively, these results demonstrate that power, entropy, and sparsity capture distinct but consistent aspects of altitude-dependent spectral evolution and motivate the unified modeling approach adopted in the ADSSM.

\subsection{ADSSM Fitting Accuracy}
\begin{figure*}[!t]
    \centering
    \subfloat[FM broadcast]{
        \includegraphics[width=0.45\textwidth]{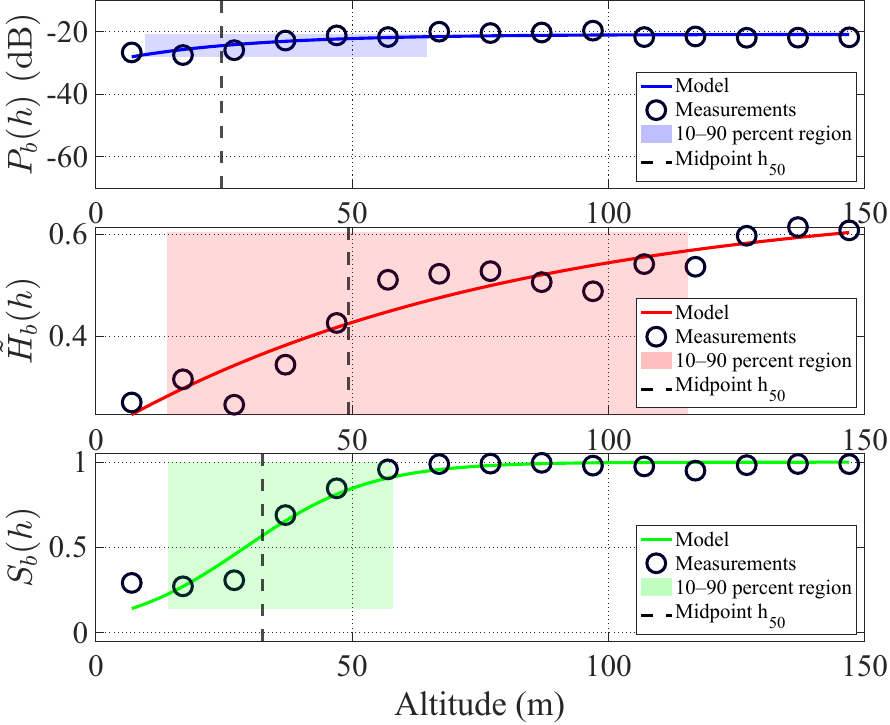}
        \label{fig:F4_TransitionRegion_2025_FM}}\vspace{-0.02in}
    \subfloat[5G NR n71 downlink]{
        \includegraphics[width=0.45\textwidth]{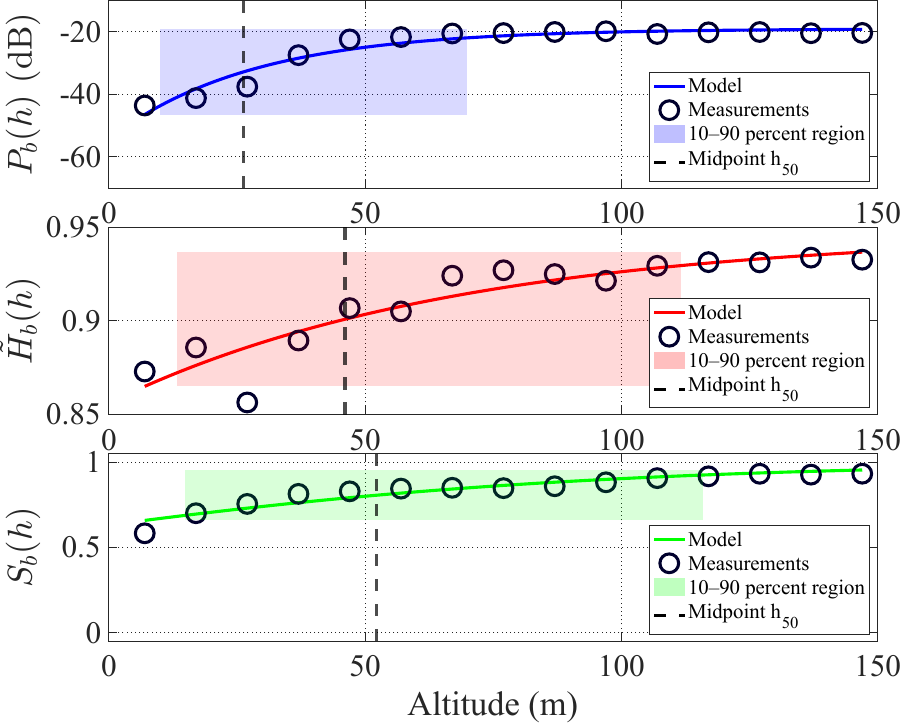}
        \label{fig:F4_TransitionRegion_2025_5G_Band_n71_DL}}\vspace{-0.02in}\\  
    \subfloat[LTE Band~13 downlink]{
        \includegraphics[width=0.45\textwidth]{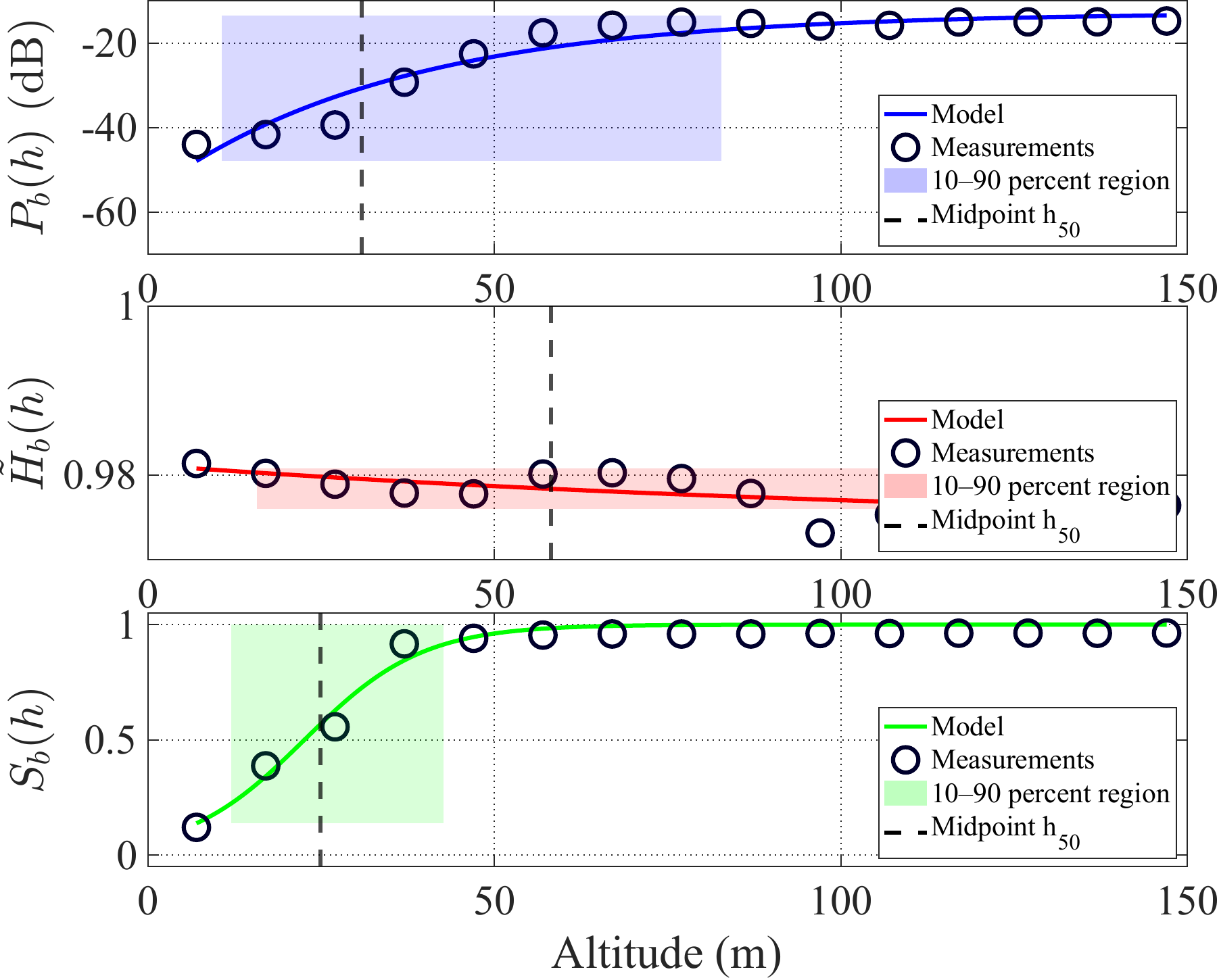}
        \label{fig:F4_TransitionRegion_2025_LTE_Band_13_DL}}\vspace{-0.02in}
    \subfloat[ISM]{
        \includegraphics[width=0.45\textwidth]{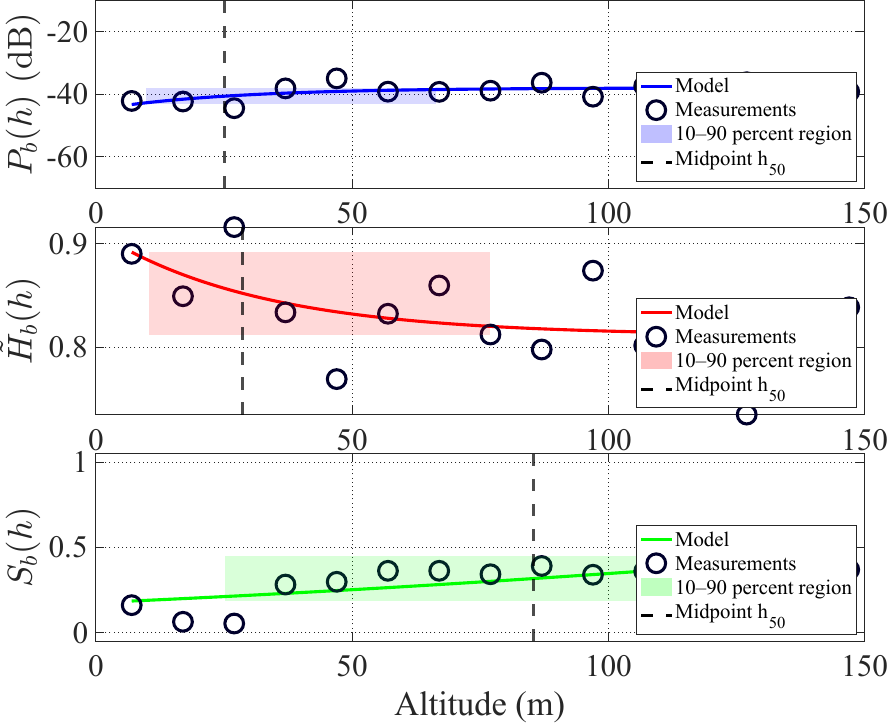}
        \label{fig:F4_TransitionRegion_2025_ISM}}\vspace{-0.02in}\\
    \subfloat[CBRS]{%
        \includegraphics[width=0.45\textwidth]{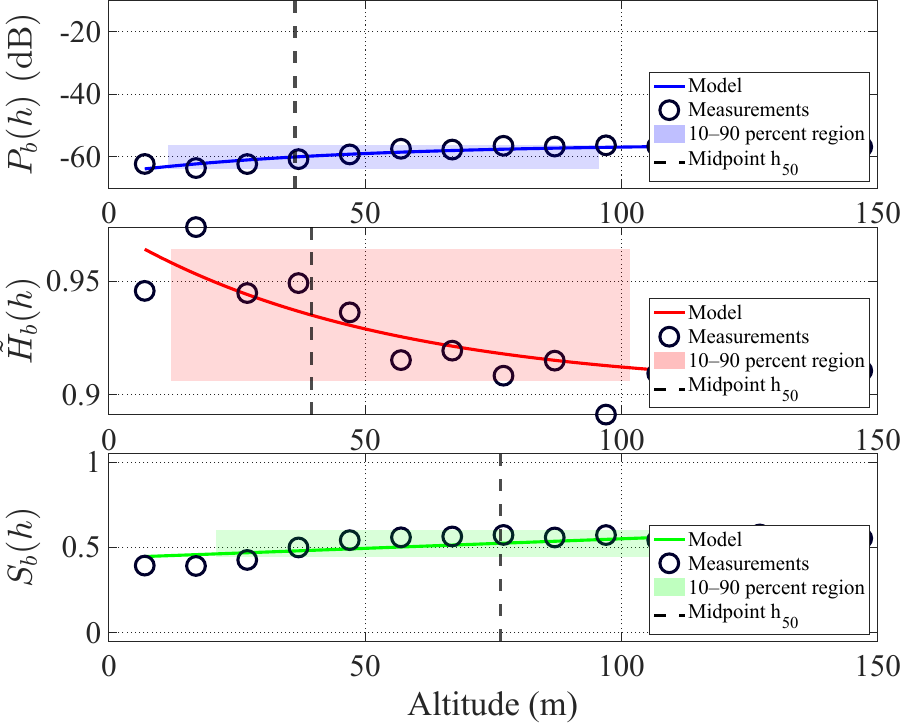}
        \label{fig:F4_TransitionRegion_2025_CBRS}}\vspace{-0.02in}
    \subfloat[5G NR C-Band]{%
        \includegraphics[width=0.45\textwidth]{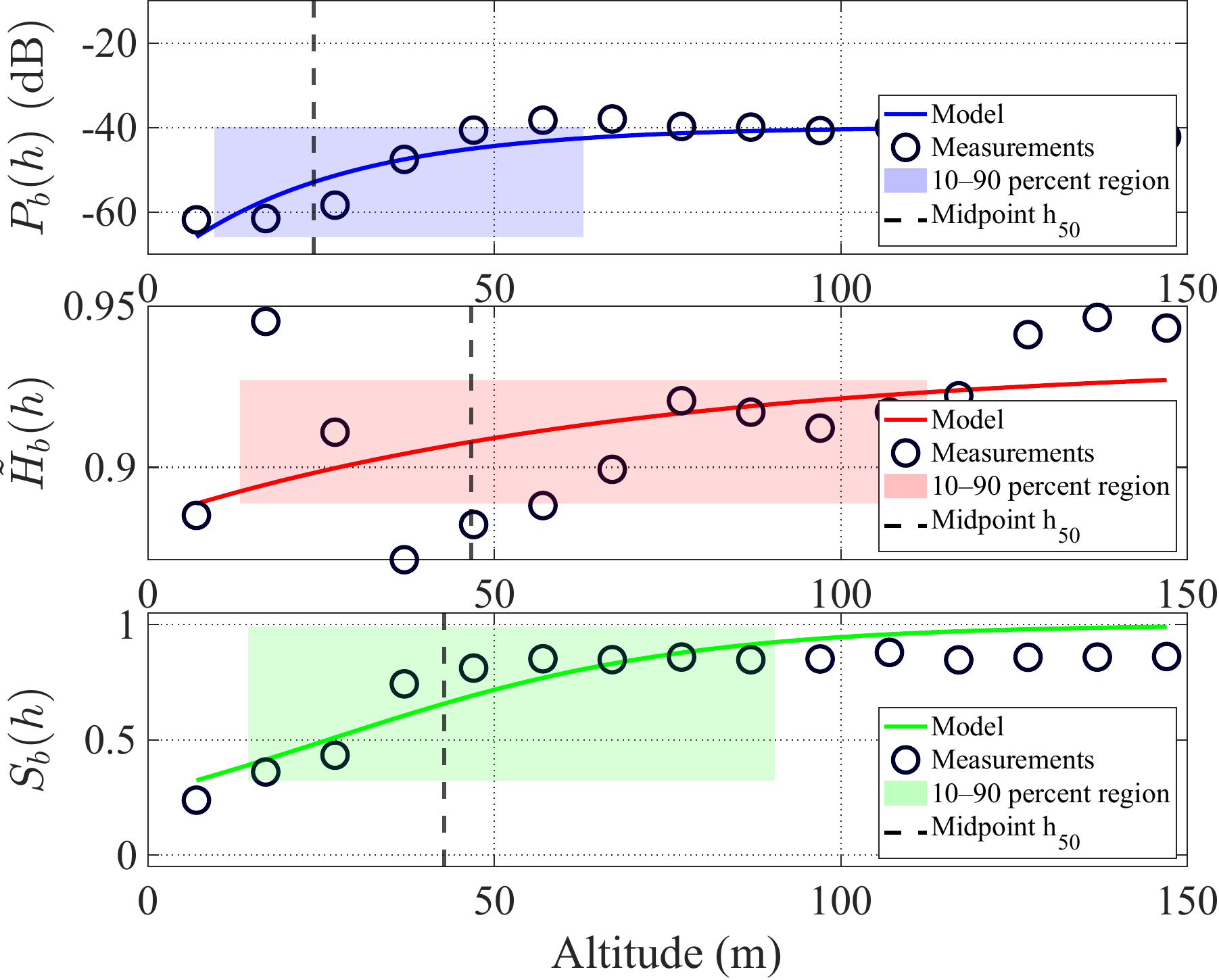}
        \label{fig:F4_TransitionRegion_2025_C-Band}}\vspace{-0.02in}
    \caption{Transition-region characterization for six measured bands based on the ADSSM power, normalized entropy, and sparsity models:
    (a) FM broadcast, (b) 5G NR n71 downlink, (c) LTE Band~13 downlink,
    (d) ISM, (e) CBRS, and (f) 5G NR C-band. Each panel displays model fits, measurements,
    the 10--90\% transition region, and the midpoint height $h_{50}$.}
    \label{fig:F4_transition_regions_allbands}
\end{figure*}

Fig.~\ref{fig:F4_transition_regions_allbands} illustrates the transition-region characterization for measured bands under consideration based on the ADSSM power, entropy, and sparsity models. 
Fig.~\ref{fig:F4_transition_regions_allbands} highlights that the altitude ranges over which transitions occur differ substantially across metrics and bands, a behavior accurately captured by the ADSSM. In the FM broadcast band, band-average power reaches its asymptotic level at relatively low altitudes, while entropy continues to evolve over a much wider altitude range, indicating the progressive visibility of additional narrowband carriers even after power saturation. Sparsity, by contrast, exhibits a clear sigmoidal transition with an early midpoint height, reflecting rapid densification of occupied channels. For the cellular downlink bands, power transitions occur predominantly at low altitudes and stabilize quickly, whereas entropy and sparsity evolve more gradually. In LTE Band~13, sparsity saturates at lower altitudes than entropy, indicating that most of the band becomes occupied relatively early as altitude increases, while the internal spectral distribution of power continues to evolve more gradually. This behavior indicates a downlink regime in which channel occupancy is established upon the restoration of LoS conditions, while finer spectral features converge over a wider range of altitudes. The ISM and CBRS bands show weaker and more diffuse transitions across all three metrics, with higher midpoint heights and broader 10--90\% regions, reflecting heterogeneous and partially utilized deployments. In the 5G NR C-band, power exhibits an early transition, while entropy and sparsity continue to increase over a broader altitude interval, indicating ongoing refinement of spectral structure beyond the main power gain. These results demonstrate that the ADSSM captures metric-specific transition behavior and yields physically meaningful transition heights that differ across bands and spectral characteristics.

\begin{table*}[!t]
\centering
\caption{Core ADSSM Parameters and Model Error Metrics for the Measured Spectrum Bands (2025)}\vspace{-3mm}
\scalebox{0.9}{
\begin{tabular}{lcccccccccc}
\toprule
\textbf{Band} &
$\mathbf{P_{b}(\infty)}$ (dB) &
$\mathbf{P_{b}(0)}$ (dB) &
$\mathbf{H_{b}(\infty)}$ (bits) &
$\mathbf{H_{b}(0)}$ (bits) &
$\mathbf{S_{b}(\infty)}$ &
$\mathbf{S_{b}(0)}$ &
\textbf{RMSE$_P$} (dB) &
\textbf{RMSE$_H$} (bits) &
\textbf{RMSE$_S$} &
$\mathbf{R^2_P}$ \\
\midrule
5G Band n71 DL
& -19.04 & -54.26 & 8.70 & 7.87 & 1.00 & 0.15 & 1.21 & 0.18 & 0.10 & 0.90 \\

LTE Band~13 DL
& -12.80 & -55.66 & 7.19 & 7.24 & 1.00 & 0.12 & 0.88 & 0.05 & 0.07 & 0.97 \\

ISM
& -37.93 & -44.90 & 7.11 & 7.99 & 1.00 & 0.16 & 1.56 & 0.29 & 0.13 & 0.57 \\

FM
& -20.84 & -30.30 & 5.67 & 1.73 & 1.00 & 0.29 & 0.99 & 0.05 & 0.08 & 0.95 \\

CBRS
& -55.88 & -65.22 & 10.18 & 10.99 & 1.00 & 0.39 & 1.74 & 0.32 & 0.14 & 0.58 \\

5G NR C-Band
& -39.86 & -74.52 & 11.37 & 10.77 & 1.00 & 0.27 & 1.70 & 0.30 & 0.13 & 0.78 \\
\bottomrule
\end{tabular}
}
\label{tab:ADSSM_core_parameters}
\end{table*}

\subsection{Core Parameters and Practical Insights}

Table~\ref{tab:ADSSM_core_parameters} reports the fitted ADSSM parameters and associated error metrics for all measured bands in the 2025 campaign and highlights clear differences in altitude-dependent behavior across spectrum allocations. For the licensed cellular bands, namely 5G NR n71 and LTE Band~13, the gap between $P_0$ and $P_{\infty}$ is large, indicating substantial power gains as altitude increases, while the corresponding RMSE$_P$ values remain below 1.3~dB and the $R^2_P$ values are high. This confirms that the first-order power model captures the dominant clutter-to-LoS transition effectively. In contrast, the ISM, CBRS, and C-band allocations exhibit smaller absolute power gains and higher fitting errors, reflecting lower transmit powers, heterogeneous deployments, and weaker altitude dependence.

The entropy parameters further distinguish cellular signals from broadcast and unlicensed bands. For LTE Band~13, $H_0$ and $H_{\infty}$ are nearly equal, indicating that the spectral structure is already well established at low altitude, whereas FM broadcast shows a large increase in entropy with altitude as additional narrowband carriers become visible. Higher-frequency bands, particularly CBRS and C-band, exhibit elevated entropy levels across all altitudes, consistent with dense and fragmented spectral occupancy. The corresponding RMSE$_H$ values remain modest, indicating that the exponential entropy model captures the observed trends despite band-specific differences.

Sparsity parameters reveal that all bands converge toward near-complete occupancy at high altitude, with $S_{\infty}$ approaching unity, while $S_0$ varies substantially across bands. Licensed cellular bands show low $S_0$ values, reflecting limited visibility of active resources at low altitude, whereas FM and CBRS exhibit higher initial sparsity due to persistent high-power transmitters and partial utilization, respectively. Across all bands, RMSE$_S$ remains low, supporting the suitability of the logistic sparsity model. Overall, the results demonstrate that the ADSSM yields physically interpretable parameters and achieves accurate fits across diverse spectrum allocations while preserving consistent behavior at both low and high altitudes.

\subsection{Limitations and Generalization}
While the ADSSM is validated using multi-year measurements, all campaigns were conducted at a single site under a consistent event setting. As a result, the fitted parameters reflect the specific propagation environment, infrastructure density, and base-station deployment of the measurement area. In particular, transition parameters such as the clutter height $h_c$, entropy stabilization height $h_e$, and sparsity midpoint $h_s$ are environment dependent and are expected to scale with local clutter statistics, base-station heights, and deployment geometry. Lower transition heights are anticipated in open rural environments, whereas higher values are expected in dense suburban or urban settings with taller structures and vegetation.
In contrast, the functional form of the ADSSM and the relative ordering of altitude-dependent transitions across power, entropy, and sparsity are expected to remain consistent across environments, as they are driven by fundamental LoS emergence and interference aggregation mechanisms rather than site-specific deployment details.

It is worth noting that all measurement datasets were calibrated to mitigate receiver-side offsets using free-space path loss analysis of FM broadcast signals. The reported power levels therefore represent calibrated received power at the helikite-mounted SDR and do not correspond to equivalent isotropically radiated power~(EIRP), as transmitter-side parameters and propagation losses are unavailable; interested readers are referred to~\cite{maeng2025altitude} for further details.

\section{Conclusion and Discussion}
A spectrum band is appropriate for sharing only if a secondary user can reliably identify idle resources, predict interference well enough to avoid harmful impact, and exploit those opportunities in a sustained and repeatable manner. Satisfying these conditions requires a multidimensional characterization of the interference environment. Aggregated received power provides a first-order measure of how severe interference is and therefore whether operation is even viable. However, power alone is insufficient for sharing decisions, since bands with comparable average power can exhibit fundamentally different internal structures, and low average power does not imply usability if interference is fragmented or unstable. As a result, power must be interpreted together with metrics that describe how interference is distributed across frequency and how predictable it is.

Spectral sparsity and spectral entropy supply this missing information. Sparsity directly quantifies resource availability and is the primary indicator of sharing feasibility, as it captures the fraction of frequency bins that are effectively idle. High sparsity implies abundant transmission opportunities and low collision risk, whereas low sparsity fundamentally limits sharing regardless of average power or predictability. Spectral entropy complements sparsity by characterizing interference structure and predictability. Low entropy indicates dominance by a small number of structured emitters that are easier to sense, model, and avoid, while high entropy reflects many comparable contributors and necessitates conservative access strategies. A share-friendly band therefore exhibits the joint condition of high sparsity, low to moderate entropy, and low to moderate aggregated power, indicating both the presence of idle resources and the ability to exploit them efficiently.

This characterization is particularly relevant for UAVs, which are expected to rely heavily on existing spectrum, including licensed cellular bands and unlicensed ISM bands, for command, control, data links, and electronic conspicuity. As highlighted in recent U.S. Department of Transportation guidance on electronic conspicuity, leveraging existing terrestrial communication infrastructure is a key enabler for safe UAV integration at scale~\cite{dot2024electronicconspicuity}. For UAVs, understanding how aggregated interference power evolves with altitude is essential for quantifying the interference environment experienced in flight, since increasing altitude fundamentally alters propagation and LoS conditions relative to ground users. At the same time, sparsity and entropy provide critical insight into how much spectrum is actually available and how reliably it can be accessed. Together, these metrics enable altitude-aware scheduling and band selection for UAVs, supporting opportunistic yet interference-safe operation that cannot be achieved using power-only or threshold-based models.

This paper introduced ADSSM, a measurement-based framework for characterizing how band-average power, spectral entropy, and sparsity evolve with altitude for signals observed by UAV platforms. The model combines first-order differential equations for power and entropy with a logistic activation model for sparsity, and is physically grounded in the transition from clutter-limited propagation at low altitudes to LoS-dominated conditions as altitude increases.
Using multi-year spectrum measurements collected by a tethered UAV across six sub-6~GHz licensed and unlicensed U.S. spectrum allocations, we demonstrated that the ADSSM accurately fits altitude-binned spectral metrics, achieving low root-mean-square error and high coefficients of determination while preserving physically consistent asymptotic behavior. The resulting parameters provide compact and interpretable descriptors of effective clutter height, asymptotic received power, entropy stabilization, and sparsity transition altitudes. The analysis highlights that these transition parameters are environment dependent and should be expected to vary with deployment geometry, terrain, and infrastructure density.
Beyond fitting accuracy, the results reveal that altitude-dependent spectrum behavior is inherently multi-dimensional. In several bands, dominant power transitions occur over narrower altitude ranges than those associated with entropy and sparsity, indicating that interference structure and spectrum occupancy may continue to evolve even after average received power has saturated. This observation underscores the limitations of power-only air-to-ground models for spectrum-aware UAV operation.

\bibliographystyle{IEEEtran}
\bibliography{references}

\end{document}